\documentclass{article}

\usepackage{arxiv}

\usepackage[utf8]{inputenc} 
\usepackage[T1]{fontenc}    
\usepackage{hyperref}       
\usepackage{url}            
\usepackage{booktabs}       
\usepackage{amsfonts}       
\usepackage{nicefrac}       
\usepackage{microtype}      
\usepackage{lipsum}		
\usepackage{graphicx}
\usepackage{natbib}
\usepackage{doi}

\usepackage{xcolor}
\usepackage{amsmath}
\usepackage{amsthm}
\usepackage{amssymb}
\usepackage{fancyvrb}
\usepackage{graphicx}
\usepackage{mathtools}
\usepackage{stmaryrd}
\usepackage{multirow}

\makeatletter
\newcommand*{\addFileDependency}[1]{
  \typeout{(#1)}
  \@addtofilelist{#1}
  \IfFileExists{#1}{}{\typeout{No file #1.}}
}
\makeatother

\newcommand\Tstrut{\rule{0pt}{2.4ex}}         
\newcommand\Bstrut{\rule[-1.2ex]{0pt}{0ex}}   
\usepackage{hyperref}
\usepackage{tikz-cd}
\usepackage{tikz}
\usetikzlibrary{arrows,positioning,patterns,shapes,decorations,bending,chains}

\newtheorem{example}{Example}

\newtheorem{definition}{Definition}
\newtheorem{prop}{Proposition}

\newcommand{\bd}[1]{\mathbf{#1}}
\newcommand{\bs}[1]{\boldsymbol{#1}}
\newcommand{\bth}{\bs{\theta}}

\DeclareMathSymbol{\mh}{\mathord}{operators}{`\-}

\newcommand{\dom}[1]{\operatorname{dom}[#1]}
\newcommand{\scmdo}{\operatorname{do}}

\newcommand{\scmbase}{\mathcal{M}}
\newcommand{\scmabst}{\mathcal{M}^\prime}

\newcommand{\intervsetbase}{\mathcal{I}}
\newcommand{\intervsetabst}{\mathcal{I}^\prime}
\newcommand{\structfuncset}{\mathcal{F}}
\newcommand{\scmdag}{\mathcal{G}}
\newcommand{\scmfamily}{\mathfrak{M}}
\newcommand{\id}{\rm id}
\newcommand{\intrange}[2]{\llbracket{#1, #2}\rrbracket}

\newcommand{\tauomega}{\tau\mh\omega}

\newcommand{\prob}{\mathbb{P}}
\newcommand{\expval}{\mathbb{E}}

\newcommand{\scm}[1]{{\texttt{SCM}#1}}
\newcommand{\dirag}[1]{{\texttt{DAG}#1}}

\newcommand{\abm}[1]{{\texttt{ABM}#1}}
\newcommand{\ode}[1]{{\texttt{ODE}#1}}
\newcommand{\lnode}[1]{{\texttt{LNODE}#1}}
\newcommand{\lode}[1]{{\texttt{LODE}#1}}
\newcommand{\rnn}[1]{{\texttt{RNN}#1}}
\newcommand{\odernn}[1]{\ode{-}\rnn{#1}}
\newcommand{\lodernn}[1]{\texttt{L}\odernn{#1}}
\newcommand{\lrnn}[1]{\texttt{L}\rnn{#1}}

\newcommand{\scminterv}[2]{\scmbase^{#1}_{#2}}

\DeclareMathOperator*{\argmin}{arg\,min}

\newcommand{\ce}[1]{\texttt{CE}_{#1}}

\newcommand{\mytitle}{Interventionally Consistent Surrogates for Agent-based Simulators}

\title{\mytitle}

\author{ 
    Joel Dyer\\University of Oxford\\
    \And Nicholas Bishop\\University of Oxford\\
    \And Yorgos Felekis\\University of Warwick\\
    \And Fabio Massimo Zennaro\\University of Bergen 
    \AND 
    Anisoara Calinescu\\University of Oxford\\
    \And Theodoros Damoulas\\University of Warwick\\
    \And Michael Wooldridge\\University of Oxford
}

\renewcommand{\headeright}{}
\renewcommand{\undertitle}{Preprint}
\renewcommand{\shorttitle}{Interventionally Consistent Surrogates for Agent-based Simulators}

\hypersetup{
pdftitle={},
pdfsubject={},
pdfauthor={J. Dyer, N. Bishop, Y. Felekis, F.M. Zennaro, A. Calinescu, T. Damoulas, M. Wooldridge},
pdfkeywords={Agent-based models, causal abstraction, policy interventions, surrogate models},
}

\begin{document}
\maketitle

\begin{abstract}
  Agent-based simulators provide granular representations of complex intelligent systems by directly modelling the interactions of the system's constituent agents. Their high-fidelity nature enables hyper-local policy evaluation and testing of \textit{what-if} scenarios, but is associated with large computational costs that inhibits their widespread use. Surrogate models can address these computational limitations, but they must behave consistently with the agent-based model under policy interventions of interest. In this paper, we capitalise on recent developments on causal abstractions to develop a framework for learning interventionally consistent surrogate models for agent-based simulators. Our proposed approach facilitates rapid experimentation with policy interventions in complex systems, while inducing surrogates to behave consistently with high probability with respect to the agent-based simulator across interventions of interest. 
  We demonstrate with empirical studies that observationally trained surrogates can misjudge the effect of interventions and misguide policymakers towards suboptimal policies, while surrogates trained for interventional consistency with our proposed method closely mimic the behaviour of an agent-based model under interventions of interest. 
\end{abstract}


\section{Introduction}

Agent-based models (\abm{s}) are a powerful tool for modelling complex decision-making systems across application domains, including the social sciences \citep{baptista2016staff}, epidemiology \citep{10.1371/journal.pcbi.1009149}, and finance \citep{cont2007volatility}. Such models provide high-fidelity and granular representations of intricate systems of autonomous, interacting, and decision-making agents by modelling the system under consideration at the level of its individual constituent actors. In this way, \abm{s} enable decision-makers to experiment with, and understand the potential consequences of, policy interventions of interest, thereby allowing for more effective control of the potentially deleterious effects that arise from the endogenous dynamics of the real-world system. In economic systems, for example, such policy interventions may take the form of imposed limits on loan-to-value ratios in housing markets as a means for attenuating housing price cycles \citep{baptista2016staff}, while in epidemiology, such interventions may take the form of (non-)pharmaceutical interventions to inhibit the transmission of a disease \citep{10.1371/journal.pcbi.1009149}. 

Whilst \abm{s} promise many benefits, their complexity generally necessitates the use of simulation studies to understand their behaviours, and their granularity can result in large computational costs even for single forward simulations. In many cases, such costs can be prohibitively large, presenting a barrier to their use as synthetic test environments for potential policy interventions in practice. Moreover, the high-fidelity data generated by \abm{s} can be difficult for policymakers to interpret and relate to policy interventions that act system-wide \citep{haldane2018interdisciplinary}. This motivates the development of simpler \emph{surrogate} models which model the underlying system at a higher level of abstraction. Such surrogates can be used in place of \abm{s} for downstream tasks where computational resources are limited. In addition, surrogates may be viewed as interpretable explanations for \abm{} behaviour, and they allow for rapid testing of population-wide interventions which may be difficult to implement or test within an \abm{}.

However, for surrogates to be useful in downstream tasks involving experimentation with possible policy interventions, they must preserve \abm{} dynamics under the external policy interventions of interest. Without imposing this condition on the constructed surrogate, there is no guarantee that the surrogate will behave similarly under external policy interventions, which in turn may lead policy-makers away from effective policies and towards suboptimal interventions. Existing methods in the literature typically apply off-the-shelf machine learning methods to learn surrogates in an observational manner, failing to account for interventional consistency.

To address this issue, we build on recent developments in \emph{causal abstraction learning} \citep{beckers2018abstracting,zennaro2023jointly}. We view the \abm{} and surrogate model as \emph{structural causal models} \citep{pearl2009causality}, and propose a framework for constructing and learning surrogate models for complex and expensive agent-based simulators that are \textit{interventionally consistent}, in the sense that they (approximately) preserve the behaviour of the \abm{} under equivalent policy interventions. This perspective enables treating the surrogate model as a causal abstraction of the \abm{} and reveals the importance of including interventional data from the \abm{} in the learning of surrogate models, given that interventions induce different distributions as,  explained by the formalism of structural causal models. We motivate our proposed methodology theoretically, and demonstrate with simulation studies that our approach permits us to learn an abstracted surrogate model for an epidemiological \abm{} that behaves consistently in multiple interventional regimes.

Our approach establishes for the first time a connection between \abm{s} and causal abstraction, allowing researchers in the field of \abm{s} to draw on the rich literature in causality for integrating causal knowledge, evaluating \emph{what-if} scenarios, and learning new abstracted models with guarantees about interventional consistency. This, in turn, will enable decision- and policy-makers to experiment with policy interventions in cheaper and more interpretable surrogate models, with assurance that the error introduced by experimenting at a higher level of abstraction is low.

\section{Background}\label{sec:back}

We first review elements of causal inference, following the framework of \citet{pearl2009causality}, and elucidate on the connection between structural causal models and \abm{s}. We also review the notion of exact transformation between structural causal models, which serves as the theoretical motivation for our framework.

\subsection{Structural Causal Models}
A structural causal model (\scm{}), is a rigorous model describing a causal system. Each component of the system is represented by a variable, whose behaviour and relation to other components is expressed by a structural equation.

\begin{definition}[\scm{s} \citep{pearl2009causality}]\label{def:scm}
A structural causal model $\scmbase$ consists of a tuple $\langle \mathbf{X},\mathbf{U},\structfuncset,\prob(\mathbf{U})\rangle $ where:
\begin{itemize}
    \setlength\itemsep{0.1em}
    \item $\mathbf{X} = \{X_i\}_{i=1}^n$, is a finite set of endogenous random variables $X_i$ each with domain $\dom{X_i}$;
    \item $\mathbf{U} = \{U_i\}_{i=1}^n$, is a finite set of exogenous random variables, each with domain $\dom{U_i}$ and each associated with an endogenous variable; 
    \item $\structfuncset = \{f_i\}_{i=1}^n$, is a finite set of measurable structural functions, one for each endogenous variable defined as
    $f_i:\dom{PA(X_i)}\times \dom{U_i}\to \dom{X_i}$,
    where $PA(X_i)\subseteq \mathbf{X} \setminus X_i$. 
    \item $\prob_{\scmbase}(\mathbf{U})$ is a joint probability distribution over the exogenous variable factorizing as $\prod_{i=1}^n \prob_{\scmbase}(U_i)$.
\end{itemize}
The model $\scmbase$ is associated with a Directed Acyclic Graph (\dirag{}) $\scmdag_{\scmbase} = \langle \mathcal{V}, \mathcal{E} \rangle$ where the set $\mathcal{V}$ of vertices  is given by $\mathbf{X} \cup \mathbf{U}$ and the set $\mathcal{E}$ of edges is given by $\{(S_j, X_i) | S_j \in PA(X_i) \cup \{U_i\}\}_{i=1}^n$.
\end{definition}

Definition \ref{def:scm} conforms to the standard definition of a \emph{Markovian \scm{}} (see Appendix \ref{app:assumptions} for an explanation of the underlying assumptions). Thanks to the measurability of the structural functions in $\mathcal{F}$, the probability distribution $\prob_{\scmbase}(\mathbf{U})$ over the exogenous variables can be pushed forward over the endogenous variables, defining the probability distribution $\prob_{\scmbase}(\mathbf{X})= \mathcal{F}_{\#}(\prob_{\scmbase}(\mathbf{U}))$. Joint probability distributions $\prob_{\scmbase}(\mathbf{S})$ can then be defined for any subset $\mathbf{S}\subseteq \mathbf{X}$. 

External interventions on the system by an experimenter can be represented in an \scm{} through changes in the structural functions. In this work, we focus our attention on hard interventions, which assign fixed values to a subset of endogenous variables.

\begin{definition}[Interventions \citep{pearl2009causality}]
Given an \scm{} $\scmbase$, $\mathbf{S}\subseteq \mathbf{X}$ and a set of values $\mathbf{s}$ realizing $\mathbf{S}$, an intervention $\iota = \scmdo(\mathbf{S}=\mathbf{s})$, is an operator that replaces each function $f_i$ associated with $S_i$ with constant $s_i$.
\end{definition}
The intervention $\iota = \scmdo(\mathbf{S}=\mathbf{s})$ induces a new \emph{post-intervention} \scm{}, $\scmbase_{\iota} = \langle \mathbf{X},\mathbf{U},\structfuncset_{\iota},\prob(\mathbf{U})\rangle $, identical to the original one, except for the set of structural functions $\structfuncset_{\iota}$ where the functions $f_i$ are replaced with the constants $s_i$. 
The probability distribution of $\scmbase_{\iota}$ is computed as $\prob_{\scmbase_{\iota}}(\mathbf{X}\setminus\mathbf{S})$. Graphically, the intervention $\iota$ mutilates the \dirag{} of $\scmbase$ by removing the incoming edges in each variable $S_i$. 

We use $\intervsetbase$ to denote a set of feasible interventions on the \scm{} $\scmbase$ that are relevant to a policymaker. Intervention sets are equipped with a natural partial ordering: let $\iota_1 = (\mathbf{S}=\mathbf{s})$ and $\iota_2 = (\mathbf{T}=\mathbf{t})$; then $\iota_1 \preceq \iota_2$ iff (i) $\mathbf{S} \subseteq \mathbf{T}$, and (ii) for each $S_i = T_i$ it holds $s_i = t_i$; informally, $\iota_1$ intervenes on a subset of the variables that $\iota_2$ intervenes on, and it sets the same values as $\iota_2$.

\subsection{Agent-based Models as \scm{}s}

Observe that an \abm{} can be modelled as a \scm{} by expressing its implicit underlying causal structure. Practically, this entails encoding quantities of interest as endogenous variables, deterministic dynamics into structural equations, and factoring sources of randomness into exogenous variables. Example \ref{ex:abm_as_scm} shows how a common \abm{} of epidemics can be described as a \scm{}.

 \begin{example}[Spatial SIRS \abm{}]\label{ex:abm_as_scm}
     We consider a susceptible-infected-recovered-susceptible (SIRS) epidemic model on an $L \times L$ lattice of cells, each of which represents one of $N = L^2$ agents. The state of each agent can be 0, 1, or 2, respectively indicating that the agent is disease-free and susceptible to infection, infected, or is recovered from a recent infection. The infection status of all agents at discrete time step $t\in \intrange{0}{T}$ is written as $\bd{x}_t \in \{0, 1, 2\}^{N}$, where $T$ is the total number of simulated time steps, and $\intrange{l}{m} = \{l,l+1,\ldots,m-1,m\}$ for integers $l \leq m$. The states $\bd{x}_{t,n}$ of each of the agents $n\in\intrange{1}{N}$ are updated synchronously as follows for $t \in \intrange{0}{T-1}$:
    \begin{itemize}    
        \setlength\itemsep{0.1em}
        \item If $\bd{x}_{t,n} = 0$, then $\bd{x}_{t+1,n} = 1$ with probability
        \begin{equation}
            p_{t,n}(\alpha_{t+1}) = 1 - (1 - \alpha_{t+1})^{\sum_{n' \in \mathcal{N}_n} \mathbb{I}[\bd{x}_{t,n'} = 1]}
        \end{equation}
        where $\mathcal{N}_n$ is the von Neumann neighbourhood for cell $n$; else remain susceptible.
                \item If $\bd{x}_{t,n} = 1$, then $\bd{x}_{t+1,n} = 2$ with probability $\beta_{t+1}$; else remain infected.
        \item If $\bd{x}_{t,n} = 2$, then $\bd{x}_{t+1,n} = 0$ with probability $\gamma_{t+1}$; else remain recovered.
    \end{itemize}
    In the above, $\bth_t = (\alpha_t, \beta_t, \gamma_t) \in [0,1]^3$ are the model parameters determining the transition probabilities between states. While these may vary over time, the simplest case consists of assigning all $\bth_t$ the same 
    vector, 
    \begin{equation}\label{eq:abm_se_3}
        \bth_t = \bd{v}\quad \quad  \forall t\in \intrange{1}{T}.
    \end{equation}
    The model is initialised by infecting each agent in the model at initial time $t=0$ with probability $I_0 \in [0,1]$. The value of $I_0$ for any forward simulation of the model can be chosen by drawing a random variable $a$ from some distribution on $[0,1]$ and setting 
    \begin{equation}\label{eq:abm_se_4}
        I_0 = a.
    \end{equation}
    With this model in place, lockdowns over some time period $t_{l}:t_{l} + \Delta$ of length $\Delta \geq 0$ can be modelled (crudely) by setting $\bth_{t_{l}:t_{l}+\Delta} = (0, \beta, \gamma)$ for $\beta, \gamma \in [0,1]$.

    To express this \abm{} as an \scm{}, we define the following:

    \paragraph{Endogenous variables} These consist of the variables of interest that may be set by the policymaker: $I_0$, $\{\bd{x}_{t}\}_{0\leq t\leq T}$, and $\{\bth_{t}\}_{1\leq t\leq T}$.

    \paragraph{Exogenous variables} The model as described above is initialised randomly according to $a$, $\bd{v}$, and a collection $\bd{u}_0 = (\bd{u}_{0,n})_{1 \leq n\leq N}$ of $N$ random variables distributed as $\mathcal{U}(0,1)$, the $n$th of which decides whether agent $n$ is infected at time $t=0$. Similarly, further collections $\bd{u}_t, t \in \intrange{1}{T}$ of $\mathcal{U}(0,1)$ random variables decide how each agent updates their state at each time step. Thus the exogenous variables for the model are $a$, $\bd{v}$, and the $\bd{u}_{t}$ for $t \in \intrange{0}{T}$.

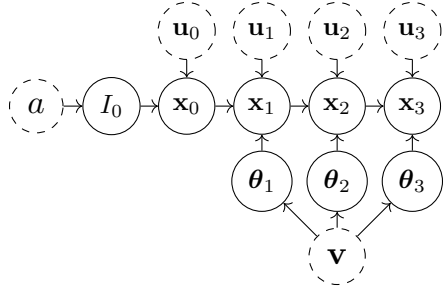
\begin{figure}[t]
\begin{center}
\begin{tikzpicture}[node distance=1cm, scale=1, ->] 
    \node[circle,draw,scale=1.25,dashed] at (0,1) (w) {$a$};
    \node[circle,draw,right of=w,] (I0) {$I_0$};
    \node[circle,draw,right of=I0,] (x0) {$\bd{x}_0$};
    \node[circle,draw,right of=x0,] (x1) {$\bd{x}_1$};
    \node[circle,draw,right of=x1,] (x2) {$\bd{x}_2$};
    \node[circle,draw,right of=x2,] (x3) {$\bd{x}_3$};
    \node[circle,draw,right of=I0,yshift=1cm,dashed] (u0) {$\bd{u}_0$};
    \node[circle,draw,right of=u0,dashed] (u1) {$\bd{u}_1$};
    \node[circle,draw,right of=u1,dashed] (u2) {$\bd{u}_2$};
    \node[circle,draw,right of=u2,dashed] (u3) {$\bd{u}_3$};
    \node[circle,draw,] at (3,0) (th1) {$\bth_1$};
    \node[circle,draw,scale=1.25,dashed] at (4,-1) (v) {$\mathbf{v}$};
    \node[circle,draw,right of=th1,] (th2) {$\bth_2$};
    \node[circle,draw,right of=th2,] (th3) {$\bth_3$};
    \draw[->] (v) -- (th1);
    \draw[->] (v) -- (th2);
    \draw[->] (v) -- (th3);
    \draw[->] (w) -- (I0);
    \draw[->] (I0) -- (x0);
    \draw[->] (x0) -- (x1);
    \draw[->] (x1) -- (x2);
    \draw[->] (x2) -- (x3);
    \draw[->] (th1) -- (x1);
    \draw[->] (th2) -- (x2);
    \draw[->] (th3) -- (x3);
    \draw[->] (u0) -- (x0);
    \draw[->] (u1) -- (x1);
    \draw[->] (u2) -- (x2);
    \draw[->] (u3) -- (x3);
\end{tikzpicture}
\end{center}
\vspace{-1em}
\caption{The directed acyclic diagram induced by the structural causal model corresponding to the spatial SIRS agent-based simulator for $T = 3$ time steps.}\label{fig:abm_dag}
\vspace{-1em}
\end{figure}

\subsection{Causal Abstractions}
Beside expressing interventions more rigorously, viewing an \abm{s} as a \scm{s} allows one to take advantage of the theory of \emph{causal abstraction} to formalise the relationship between an \abm{} and a surrogate model.
Indeed, causal abstraction provides a framework for relating \scm{s} representing an identical system at different levels of granularity. The notion of exact transformation formalizes this relation, providing a framework to relate complex models, such as \abm{s}, to simpler top-down models while preserving causal structure. 

\begin{definition}[$\tauomega$ Exact Transformation \citep{rubenstein2017causal}]\label{def:tauomega}
Given two \scm{s}, $\scmbase$ and $\scmabst$, with respective intervention sets $\intervsetbase$ and $\intervsetabst$, a \emph{$\tauomega$ transformation} is a pair $(\tau, \omega)$ consisting of a map $\tau: \dom{\mathbf{X}} \rightarrow \dom{\mathbf{X'}}$ and a surjective, order-preserving map $\omega: \intervsetbase \rightarrow \intervsetabst$. An \emph{exact} $\tauomega$ transformation is a $\tauomega$ transformation such that 
\begin{equation}
\label{eq:exacttrans}
\tau_{\#} (\prob_{\scminterv{}{\iota}})= \prob_{\scminterv{\prime}{\omega(\iota)}}, \forall \iota \in \intervsetbase.
\end{equation}
\end{definition}

    \paragraph{Structural equations} Equations \ref{eq:abm_se_3} and \ref{eq:abm_se_4}, respectively, define the structural equations $f_{\bth_t}$ and $f_{I_0}$ for the endogenous variables $\bth_t$ and $I_0$. The structural equation $f_{\bd{x}_{0,n}}$ for each $\bd{x}_{0,n}, n\in\intrange{1}{N}$ can furthermore be written as
    \begin{equation}\label{eq:abm_se_2}
        \bd{x}_{0,n} = f_{\bd{x}_{0,n}}(u_{0,n}, I_0) = \mathbb{I}\left[u_{0,n} < I_0\right].
    \end{equation}
    Finally, the three update rules for $\bd{x}_t$ described above can be written concisely with the following structural equations for $t \in \intrange{0}{T-1}$:
    \begin{align}\nonumber 
        \bd{x}_{t+1,n} &= f_{\bd{x}_{t+1,n}}(\bth_{t+1}, u_{t+1,n}, \bd{x}_{t,n})\\\label{eq:abm_se_1}
        &= \mathbb{I}\left[\bd{x}_{t,n} = 0\right]\cdot \mathbb{I}\left[u_{t+1,n} < p_{t,n}(\alpha_{t+1})\right] + \mathbb{I}\left[\bd{x}_{t,n} = 1\right]\cdot (1 + \mathbb{I}\left[u_{t+1,n} < \beta_{t+1}\right])\\\nonumber
                       &\qquad \qquad \qquad + 2\mathbb{I}\left[\bd{x}_{t,n} = 2\right]\cdot (1 - \mathbb{I}\left[u_{t+1,n} < \gamma_{t+1}\right]),
    \end{align}

    \paragraph{Distribution over the exogenous variables} The stochastic behaviour over the exogenous variables is fully specified by the distribution over $a$ and $\bd{v}$, together with a $\mathcal{U}(0,1)$ distribution over each $\bd{u}_t$.

    \paragraph{The underlying graph} The \dirag{} corresponding to this \scm{} is shown in Figure \ref{fig:abm_dag} for $T=3$. 
    
    In this model, interventions in the form of, for example, a lockdown can be (crudely) modelled by intervening on one or more of the $\bth_t$ as $\scmdo(\bth_t = (0, \beta, \gamma))$ for some $\beta, \gamma \in [0,1]$, while in the observational regime the $\bth_t$ will all be assigned the same value. 
\end{example}

\begin{figure}[t]
\centering
\begin{tikzcd}[row sep=huge,column sep=huge]
\iota \arrow{rr}{\scmbase} \arrow[swap]{d}{\omega} & & \prob_{\scminterv{}{\iota}} \arrow{d}{\tau} \\
\omega(\iota) \arrow{rr}{\scmabst} &  & \prob_{\scminterv{\prime}{\omega(\iota)}} 
\end{tikzcd}
\caption{Computing $\tau_{\#} (\prob_{\scminterv{}{\iota}})$, corresponds to moving right then down in the diagram. That is, running the intervention $\iota$ in a base model $\scmbase$ such as an \abm{}. Likewise, computing $\prob_{\scminterv{\prime}{\omega(\iota)}}$ corresponds to moving down then right. That is, running the intervention $\omega(\iota)$ in an abstracted model $\scmbase'$ such as a surrogate. If $(\tau, \omega)$ is an exact transformation, then the diagram is commutative for all interventions. That is, $\tau_{\#} (\prob_{\scminterv{}{\iota}}) = \prob_{\scminterv{\prime}{\omega(\iota)}}$ for all $\iota \in \intervsetbase$. 
}\label{fig:exact}
\vspace{-1em}
\end{figure}
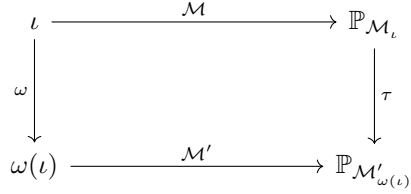

An exact $\tauomega$ transformation constitutes a form of abstraction between probabilistic causal models \citep{beckers2020approximate} with the guarantee of commutativity between intervention and transformation as detailed in Figure \ref{fig:exact}: intervening via $\iota$ and then abstracting produces the same result as abstracting first and then intervening via $\omega(\iota)$. The map $\tau$ describes corresponding states in each of the models, while the map $\omega$ describes corresponding interventions in each model.
In the following, whenever the map $\tau$ is clear from context, we shorthand the pushforward measure $\tau_{\#} (\prob_{\scminterv{}{\iota}})$ as $\prob_{\scmbase_\iota}^{\#}$.

An exact $\tauomega$ transformation between the \scm{} $\scmbase$ underlying an \abm{} and the \scm{} $\scmabst$ underlying the candidate surrogate model would (a) certify that the surrogate preserves the causal structure of interest, (b) allow to interpret the emergent causal structure of the \abm{} through $\scmabst$, and (c) provide guarantees of interventional consistency when a policymaker would study real-world interventions through the surrogate model.

\section{Abstraction Error}
\label{sec:abserror}
Unfortunately, it is unrealistic to assume that an exact $\tauomega$ transformation exists between an \abm{} and a surrogate. In fact, \abm{s} are often developed for problem settings where simple top-down 
models fail to fully capture the dynamics of interest. In such cases, a more pragmatic goal is 
to find an approximate abstraction \citep{beckers2020approximate} from the \abm{} to the surrogate. For this reason we define the \emph{abstraction error}.

\begin{definition}[Abstraction error]\label{def:abstractionerr}
Let $(\tau,\omega)$ be a $\tauomega$ transformation between two \scm{s} $\scmbase$ and $\scmabst$ with respective intervention sets $\intervsetbase$ and $\intervsetabst$. Given a statistical divergence $d$ between distributions, and a distribution $\eta$ over the intervention set $\intervsetbase$, we define the \emph{abstraction error} as follows:
\begin{equation}
d_{\tau, \omega}(\scmbase, \scmabst) = \expval_{\iota \sim \eta} \left[ d\left(\tau_{\#} (\prob_{\scminterv{}{\iota}}) ,\, \prob_{\scminterv{\prime}{\omega(\iota)}}\right)\right].
\end{equation}
A $\tauomega$ transformation is \emph{$\alpha$-approximate} for some $\alpha \in \mathbb{R}_{\geq 0}$
if $d_{\tau, \omega}(\scmbase, \scmabst) \leq \alpha$.
\end{definition}

\begin{figure}[t]
\centering
\begin{tikzcd}[row sep=normal,column sep=large]
\iota \arrow{rr}{\scmbase} \arrow[swap]{dd}{\omega} & & \prob_{\scminterv{}{\iota}} \arrow{d}{\tau} \\
 & & \tau_{\#} (\prob_{\scminterv{}{\iota}}) \\
\omega(\iota) \arrow{r}{\scmabst} & \prob_{\scminterv{\prime}{\omega(\iota)}} \arrow[ru, dashed, leftrightarrow, red, "d_{\tau, \omega}\leq \alpha"]& 
\end{tikzcd}
    \caption{When computing abstraction error, we compare the distributions $\tau_{\#} (\prob_{\scminterv{}{\iota}})$ and $\prob_{\scminterv{\prime}{\omega(\iota)}}$ for each intervention $\iota$ using the divergence $d_{\tau, \omega}$, as indicated by the red dotted arrow. If the divergence is zero then we recover the commutative diagram in Figure \ref{fig:exact}.}\label{fig:err}
\end{figure}
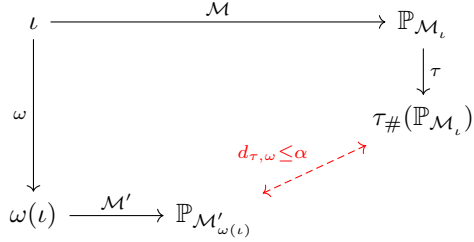

The distributional distance $d$ describes the similarity between distributions over states of the surrogate model. A $\tauomega$ abstraction with low abstraction error implies that $\tau_{\#}(\prob_{\scminterv{}{\iota}})$ is close to $\prob_{\scminterv{\prime}{\omega(\iota)}}$ in expectation with respect to the interventional distribution $\eta$. If the $d\left(\tau_{\#} (\prob_{\scminterv{}{\iota}}) ,\, \prob_{\scminterv{\prime}{\omega(\iota)}}\right)$ is zero for all interventions $\iota \in \intervsetbase$, then $(\tau, \omega)$ is an exact transformation. Abstraction error is summarised pictorially in Figure \ref{fig:err}.

Definition \ref{def:abstractionerr} differs from previously defined notions of abstraction error in the causal abstraction literature. Whilst \cite{beckers2020approximate} employ a maximum over interventions, we instead take an expectation over a fixed interventional distribution $\eta$. This is motivated by the fact that policymakers will often hold prior preferences over possible interventions, which may, for example, reflect the cost of implementing each intervention in the real world. Through the specification of $\eta$, one may implicitly favour surrogates which perform well with respect to interventions of high interest. Further discussion is provided in Appendix \ref{app:details}.

\section{Method}
\label{sec:method}
The definitions of abstraction and abstraction error provide us with a grounded framework for learning surrogates, and, in the remainder, we assume that the base model $\scmbase$ is always implicitly represented by an \abm. Our goal is to identify a surrogate model which is interventionally consistent with a given \abm{}. Specifically, given a set of candidate surrogate models $\scmfamily$, we aim to identify a surrogate model and a $\tauomega$ transformation that minimises the abstraction error. 

To proceed, we assume that $\scmfamily$ is induced by a parameterised family $\{\scmbase^{\psi} \: : \: \psi \in \Psi\}$ of differentiable surrogate simulators with tractable probability mass or density function $q^{\psi}$. Here, $\scmbase^{\psi}$ denotes the causal model induced by a surrogate with parameters $\psi$, and $\Psi$ denotes the set of feasible parameter values. When clear from context, we refer to each surrogate by their corresponding parameter value $\psi \in \Psi$. Such a family of surrogate models can be constructed through a composition of differential equation- or deep learning-based modelling, in combination with probability distributions with reparameterisable sampling procedures; an example of such a composition, of which we make use in the experiments presented in Section \ref{sec:exp}, is a latent neural ordinary differential equation model \citep{rubanova2019latent}. We further assume only the ability to sample from $\tau_{\#} (\prob_{\scmbase})$, which amounts to running the \abm{} and applying $\tau$ to the output. 

Generally speaking, policymakers know what macroscopic quantities are of interest when modelling a complex system, and how to aggregate the microscopic variables into global statistics. For example, in macroeconomic settings, policymakers will often be concerned with aggregate quantities such as unemployment rates or aggregate demand, which can be derived from the state of the agents. We thus assume access to a pre-specified map $\tau$ defining the aggregate, emergent quantities of interest to the policymaker.

Hence, to find an appropriate $\tauomega$ transformation, we need only identify an intervention map $\omega^{\star}$ between $\intervsetbase$ and $\intervsetabst$. To limit computational burden, we select $\omega^{\star}$ from a parameterised family $\Omega \vcentcolon= \{\omega^{\phi} \: : \: \phi \in \Phi\}$ with parameters $\phi$ ranging over the set $\Phi$. For example, $\phi$ may correspond to the weights of a neural network. As with surrogates, we refer to each feasible intervention map via their corresponding parameter value $\phi \in \Phi$.

We then select $\phi^{\star}$ and $\psi^\star$ jointly by minimising $d_{\tau,\omega}(\scmbase, \scmbase')$ over $\Omega \times \scmfamily$:
\begin{equation*}
    \phi^{\star}, \psi^{\star}
    = 
    \argmin_{\phi \in \Phi, \psi \in \Psi}d_{\tau, \omega^{\phi}}(\scmbase, \scmbase^{\psi}).
\end{equation*}
Since each element of $\scmfamily$  has a differentiable and tractable distribution, a convenient choice of discrepancy $d$ is the Kullback-Leibler (KL) divergence:
\begin{equation}
    d\bigg(\tau_{\#} (\prob_{\scmbase_{\iota}}), \prob_{\scmbase^{\psi}_{\omega^{\phi}(\iota)}}\bigg) = \mathbb{E}_{\prob_{\scmbase_{\iota}}^{\#}} \left[ \log \frac{{\rm d} \prob^{\#}_{\scmbase_{\iota}} }{ {\rm d} \prob_{\scmbase_{\omega^{\phi}(\iota)}^{\psi}} } \right].
\end{equation}
The KL divergence can be minimised with gradient assisted optimisation procedures, in which a Monte Carlo estimate of the gradient is obtained as
\begin{equation*}
    \nabla_{\phi,\psi}\, d_{\tau,\omega^{\phi}}(\scmbase, \scmbase^{\psi}) \approx \frac{1}{B} \sum_{b=1}^B -\nabla_{\phi,\psi}
    \log
    q^{\psi}_{\omega^{\phi}(\iota^{(b)})}
    (\mathbf{y}^{(b)})
\end{equation*}
where $\iota^{(b)} \sim \eta$, $\bd{y}^{(b)} \sim \tau_{\#}( \prob_{\scmbase_{\iota^{(b)}}} )$, $q^{\psi}_{\omega^{\phi}(\iota)}$ is the probability mass or density function associated with $\scmbase^{\psi}_{\omega^{\phi}(\iota)}$, and $B \geq 1$ is the size of a batch drawn from $R \geq B$ training examples from the joint distribution over the $\iota^{(b)}$ and $\bd{y}^{(b)}$. After the pair $(\phi^{\star}, \psi^{\star})$ have been selected, we may generate data from the macromodel for \abm{} intervention $\iota$ by sampling from $\prob_{\scmbase^{\psi^\star}_{\omega^{\phi^\star}(\iota)}}$. 

In practice, the experimenter may be interested in only a subset of the endogenous variables of the two \scm{s}. In this case, the abstraction error we consider can instead be defined with respect to this subset, and the likelihood evaluations required in the optimisation procedure described above can be understood to contain an implicit marginalisation over the auxiliary variables that are excluded from consideration.

\subsection{Theory}

Definition \ref{def:abstractionerr} is closely related to exact transformations:
\begin{prop}\label{prop:1}
    Fix an interventional distribution $\eta$ and let $d$ be any statistical divergence. Additionally, let $(\tau, \omega)$ be a $\tauomega$ transformation between two \scm{s} $\scmbase$ and $\scmabst$. If $\tauomega$ is 0-approximate ($d_{\tau, \omega}(\scmbase, \scmabst) = 0$), 
    then we have $\eta$-almost-surely 
    \begin{equation*}
        \tau_{\#} (\prob_{\scminterv{}{\iota}})= \prob_{\scminterv{\prime}{\omega(\iota)}}.
    \end{equation*}
\end{prop}
In particular, when $\intervsetbase$ is finite and $\eta(\iota) > 0\ \forall \iota \in \intervsetbase$, then any $0$-approximate $\tauomega$  transformation is an exact $\tauomega$ transformation between the models $\scmbase$ and $\scmabst$. Note that any $f$-divergence with strictly convex $f$, such as the KL divergence, is a statistical divergence.

Definition \ref{def:abstractionerr} employs an expectation over an interventional distribution. As a result, even when the abstraction error is low, there may still be a large discrepancy between $\tau_{\#} (\prob_{\scminterv{}{\iota}})$ and $\prob_{\scminterv{\prime}{\omega(\iota)}}$ for some fixed intervention $\iota \in \intervsetbase$. Put differently, running the surrogate with intervention $\omega(\iota)$ may be a poor proxy for running the ABM with intervention $\iota$. 

The following proposition provides an upper bound on the divergence associated with any intervention sampled from the interventional distribution $\eta$ when $d$ is the KL-divergence and the \abm{} state space is finite.

\begin{prop}\label{prop:2}
    Let $d$ be the KL divergence. Moreover, let $\ce{\iota}$ denote the cross-entropy of $\prob_{\scminterv{\prime}{\omega(\iota)}}$ with respect to $\tau_{\#} (\prob_{\scminterv{}{\iota}})$:
    \begin{equation*}
        \ce{\iota} \vcentcolon =  \mathbb{E}_{\bd{Y} \sim \tau_{\#}(\prob_{\scmbase_{\iota}})}\left[ - \log q_{\omega(\iota)} \left( \bd{Y} \right) \right].
    \end{equation*}
    Assume $\dom{\mathbf{X}}$ is finite. Then for all $\varepsilon > 0$,
    \begin{equation*}
        \prob_{\eta}\left(d\left(\tau_{\#} (\prob_{\scminterv{}{\iota}}) ,\, \prob_{\scminterv{\prime}{\omega(\iota)}}\right) \geq \varepsilon\right) \leq  \frac{\expval_{\iota \sim \eta}[\ce{\iota}]}{\varepsilon}.
    \end{equation*}
\end{prop}
Evaluating the density associated with $\tau_{\#} (\prob_{\scminterv{}{\iota}})$ and, by extension, the abstraction error is often not possible for \abm{s} due to their complexity. In contrast, the upper bound in Proposition \ref{prop:2} relies only on the cross-entropy which can be estimated with finite samples.

\section{Case Study}\label{sec:exp}

\begin{figure}[t]
\begin{center}
\begin{tikzpicture}[node distance=1cm, scale=1, ->] 
    \node[circle,draw,scale=1.25,dashed] at (0,1) (w) {$\tilde{a}$};
    \node[circle,draw,right of=w,] (I0) {$\tilde{I}_0$};
    \node[circle,draw,right of=I0,] (x0) {$\tilde{\bd{y}}_0$};
    \node[circle,draw,right of=x0,] (x1) {$\tilde{\bd{y}}_1$};
    \node[circle,draw,right of=x1,] (x2) {$\tilde{\bd{y}}_2$};
    \node[circle,draw,right of=x2,] (x3) {$\tilde{\bd{y}}_3$};
    \node[circle,draw,right of=I0,yshift=1.15cm,dashed] (u0) {$\tilde{\bd{u}}_0$};
    \node[circle,draw,right of=u0,dashed] (u1) {$\tilde{\bd{u}}_1$};
    \node[circle,draw,right of=u1,dashed] (u2) {$\tilde{\bd{u}}_2$};
    \node[circle,draw,right of=u2,dashed] (u3) {$\tilde{\bd{u}}_3$};
    \node[circle,draw,right of=x0,yshift=-1.15cm] (th1) {$\tilde{\bth}_1$};
    \node[circle,draw,right of=th1,yshift=-1.15cm,scale=1.25,dashed] (v) {$\tilde{\mathbf{v}}$};
    \node[circle,draw,right of=th1,] (th2) {$\tilde{\bth}_2$};
    \node[circle,draw,right of=th2,] (th3) {$\tilde{\bth}_3$};
    \draw[->] (v) -- (th1);
    \draw[->] (v) -- (th2);
    \draw[->] (v) -- (th3);
    \path[every node/.style={font=\sffamily\small}]
      (I0) edge[bend right] node [left] {} (x1);
    \path[every node/.style={font=\sffamily\small}]
      (I0) edge[bend right] node [left] {} (x2);
    \path[every node/.style={font=\sffamily\small}]
      (I0) edge[bend right] node [left] {} (x3);
    \draw[->] (w) -- (I0);
    \draw[->] (I0) -- (x0);
    \draw[->] (x0) -- (x1);
    \draw[->] (x1) -- (x2);
    \draw[->] (x2) -- (x3);
    \draw[->] (th1) -- (x1);
    \draw[->] (th1) -- (x2);
    \draw[->] (th1) -- (x3);
    \draw[->] (th2) -- (x2);
    \draw[->] (th2) -- (x3);
    \draw[->] (th3) -- (x3);
    \draw[->] (u0) -- (x0);
    \draw[->] (u1) -- (x1);
    \draw[->] (u2) -- (x2);
    \draw[->] (u3) -- (x3);
\end{tikzpicture}
\end{center}
\vspace{-1em}
\caption{The directed acyclic graph diagram induced by the \scm{s} corresponding to the surrogate families for $T = 3$.}\label{fig:surr_dag}
\vspace{-1em}
\end{figure}
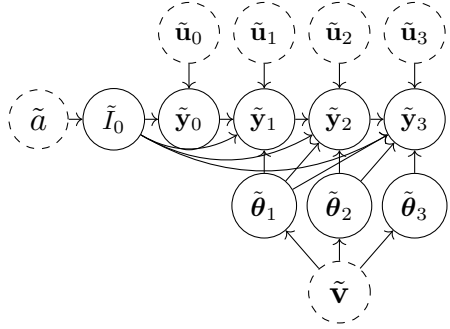

In this section, we consider a case study in which we learn interventionally consistent surrogate macromodels for the spatial SIRS \abm{} described in Example \ref{ex:abm_as_scm}. We consider three families of surrogate macromodels with endogenous variables $\tilde{I}_0 \in [0,1]$, $\tilde{\bth}_t \in \mathbb{R}_{\geq 0}^3$ for $t \in \intrange{1}{T}$, and $\tilde{\bd{y}}_t \in \{(a,b,c)\ |\ a,b,c \in \intrange{0}{N}, a + b + c = N\}$ for $t \in \intrange{0}{T}$. The \dirag{s} underlying the \scm{s} of each of these three families can be drawn as in Figure \ref{fig:surr_dag}, and the three families differ only in the form of the structural equations mapping from $\tilde{I}_0$ and $\tilde{\bth}_{0:t}$ to the $\tilde{\bd{y}}_t$. Throughout, we let $q^{\psi}$ be a Multinomial emission distribution and $\psi$ be trainable parameters of these structural equations.

\vspace{-0.7em}
\paragraph{Surrogate family 1} consists of a latent \ode{} (\lode{}) constructed by placing $q^\psi$ onto the classical SIRS \ode{}, such that the \ode{'s} three state variables (which take values in the two-simplex) index the class probabilities of $q^{\psi}$. Here, $\psi = \emptyset$.

\vspace{-0.7em}
\paragraph{Surrogate family 2} consists of a latent \odernn{} (\lodernn{}) constructed by running a recurrent neural network (\rnn{}) with trainable parameters $\psi$ over the output of the classical SIRS \ode{}. The \rnn{} output then indexes the class logits of $q^{\psi}$.

\vspace{-0.7em}
\paragraph{Surrogate family 3} consists of a latent \rnn{} (\lrnn{}) constructed by running an \rnn{} with trainable parameters $\psi$ over the $\tilde{\bth}_t$, where the output of the \rnn{} at each $t \in \intrange{0}{T}$ indexes the class logits of $q^{\psi}$.

Given $\tilde{\bth}_{1:T}, \tilde{I}_0$, these surrogate macromodels enjoy tractable likelihood functions given by
\vspace{-0.5em}
\begin{equation*}
    q^{\psi}(\tilde{\bd{y}}_{0:T} \mid \tilde{\bth}_{1:T}, \tilde{I}_0) = q^{\psi}(\tilde{\bd{y}}_0 \mid \tilde{I}_0) \prod_{t = 1}^T q^{\psi}(\tilde{\bd{y}}_t \mid \tilde{\bth}_{1:t}, \tilde{I}_0).
\end{equation*}
We provide further details in Appendix \ref{app:exp}.

\subsection{Interventions and the $\tauomega$ Transformation}

We consider two subsets of interventions at the level of the \abm{}. Denoting
\begin{align}
    \iota_{\bd{v},a} &= \scmdo\left(\bth_{1:T} = \bd{v}, I_0 = a\right),\\\nonumber
    \iota_{\bd{v},a,t_{l}} &= \scmdo\left(\bth_{1:t_{l}-1} = \bd{v}, \bth_{t_{l}:t_{l}+5} = \bd{v} \odot (0, 1, 1),\right.\\
    & \quad \quad \quad  \left. \bth_{t_{l} + 6:T} = \bd{v}, I_0 = a\right),
\end{align}
we define $\intervsetbase = \intervsetbase_{\rm init} \cup \intervsetbase_{\rm init,\, lock}$, where
\begin{align}
    \intervsetbase_{\rm init} &= \{  \iota_{\bd{v},a} \mid (\bd{v},a) \in [0,1]^4 \},\\
    \intervsetbase_{\rm init,\, lock} &= \{ \iota_{\bd{v},a,t_{l}} \mid (\bd{v},a, t_{l}) \in [0,1]^4 \times \intrange{5}{10} \}.
\end{align}
The first of these is a subset of interventions that fix the initial proportion of infected individuals in the agent-based simulator, as well as its parameter values. The second subset of interventions is the set of interventions that fix (a) the initial proportion of infected individuals in the \abm{}, (b) the values of the \abm{'s} parameters before, during, and beyond a lockdown beginning at time $t_{l} \in \intrange{5}{10}$ with duration equal to 5 time steps, and (c) the value of $t_{l}$. Similarly defining
\begin{align}
    \iota'_{\tilde{\bd{v}},\tilde{a}} &= \scmdo\left(\tilde{\bth}_{1:T} = \tilde{\bd{v}}, \tilde{I}_0 = \tilde{a}\right),\\\nonumber
    \iota'_{\tilde{\bd{v}},\tilde{a},\tilde{t}_{l}} &= \scmdo\left(\tilde{\bth}_{1:\tilde{t}_{l}-1} = \tilde{\bd{v}}, \tilde{\bth}_{\tilde{t}_{l}:\tilde{t}_{l}+5} = \tilde{\bd{v}} \odot (0, 1, 1), \right.\\
    & \quad \quad \quad  \left. \tilde{\bth}_{\tilde{t}_{l} + 6:T} = \tilde{\bd{v}}, \tilde{I}_0 = \tilde{a}\right),
\end{align}
we furthermore define the intervention set $\intervsetabst = \intervsetabst_{\rm init} \cup \intervsetabst_{\rm init,\, lock}$ for the macromodels, where
\begin{align}
    \intervsetabst_{\rm init} &= \{  \iota'_{\tilde{\bd{v}},\tilde{a}} \mid (\tilde{\bd{v}},\tilde{a}) \in  \mathbb{R}_{\geq 0}^3 \times [0,1] \},\\\nonumber
    \intervsetabst_{\rm init,\, lock} &= \{ \iota'_{\tilde{\bd{v}},\tilde{a},\tilde{t}_{l}} \mid \\
                                   & (\tilde{\bd{v}}, \tilde{a}, \tilde{t}_{l}) \in \mathbb{R}_{\geq 0}^3 \times [0,1] \times \intrange{5}{10} \}.
\end{align}
The map $\tau$ is taken to map: $\bth_{t}$ identically to $\tilde{\bth}_{t}$ for each $t \in \intrange{1}{T}$; the microstate $\bd{x}_{t}$ of the \abm{} at each time step to the $\tilde{\bd{y}}_t$ through an aggregation map that counts the number of agents in $\bd{x}_t$ in each of the three possible states (susceptible, infectious, and recovered); and the initial proportion $I_0$ of infected agents in the \abm{} identically to $\tilde{I}_0$. The map $\omega^{\phi}$ is taken to act as
\begin{equation}\label{eq:omega_obs}
    \omega^{\phi} : \quad \iota_{\bd{v},a} \mapsto \iota'_{f^{\phi}(\bd{v}),a}\quad \ ,\ \quad \iota_{\bd{v},a,t_{l}} \mapsto \iota'_{f^{\phi}(\bd{v}),a,t_{l}}
\end{equation}
for a neural network $f^\phi : [0,1]^3 \to \mathbb{R}_{\geq 0}^3$. 

\begin{table*}[t]
\caption{Metrics for interventionally ($\mathbf{I}$) and observationally ($\mathbf{O}$) trained surrogates on interventional ($\mathbf{I}'$) and observational ($\mathbf{O}'$) test set ($\text{median}_{\text{first quantile}}^{\text{third quantile}}$ from 5 repeats). Bold indicates best performance.} \label{tab:int_obs}
\begin{center}
\begin{tabular}{c|l|cc|cc|cc}
\textbf{Test} & \textbf{Model} & \multicolumn{2}{|c|}{\lrnn{}} & \multicolumn{2}{|c|}{\lodernn} & \multicolumn{2}{c}{\lode{}}\\
{} & \textbf{Train}     &\textbf{I}  &\textbf{O}  &\textbf{I}  &\textbf{O}   &\textbf{I}  &\textbf{O}\\
\hline
\multirow{2}{*}{$\mathbf{I'}$} & AMSE \Tstrut\Bstrut $(\times 10^{-1})$ & 
$\mathbf{3.475_{3.405}^{3.910}}$ & $49.44_{46.66}^{52.61}$ & 
$\mathbf{3.350_{3.179}^{3.410}}$ & $18.47_{17.12}^{21.91}$ & 
$\mathbf{8.150_{8.058}^{8.242}}$ & $22.39_{22.14}^{22.65}$ \\
& ANLL \Bstrut $(\times 10^{3})$ & 
$\mathbf{2.091_{2.030}^{2.164}}$  & $21.77_{20.13}^{22.93}$ & 
$\mathbf{1.987_{1.980}^{1.990}}$  & $8.396_{8.269}^{9.885}$  & 
$\mathbf{4.008_{3.999}^{4.017}}$ & $10.03_{9.909}^{10.14}$\\
\hline
\multirow{2}{*}{$\mathbf{O'}$ } & AMSE \Tstrut\Bstrut $(\times 10^{-1})$ & 
$4.127_{4.105}^{4.259}$ & $\mathbf{2.945_{2.623}^{3.155}}$ & 
$3.592_{3.538}^{3.681}$ & $\mathbf{2.523_{2.164}^{2.783}}$ & 
$18.42_{18.14}^{18.71}$ & $\mathbf{4.361_{4.321}^{4.401}}$ \\
& ANLL $(\times 10^{3})$ 
& $2.220_{2.156}^{2.231}$ & $\mathbf{1.641_{1.431}^{1.709}}$ 
& $1.855_{1.850}^{1.968}$  & $\mathbf{1.429_{1.267}^{1.528}}$ 
& $7.629_{7.517}^{7.741}$ & $\mathbf{2.151_{2.131}^{2.170}}$
\end{tabular}
\vspace{-1em}
\end{center}
\end{table*}

\subsection{The Benefits of Training with Interventional Data}

To explore the benefits of training a surrogate macromodel with interventional consistency in mind, we jointly learn the parameters $\phi, \psi$ of the surrogates and the map $\omega^{\phi}$ described above in two different ways\footnote{``Observationally'' and ``interventionally'' are misnomers: both are trained interventionally, but differ in the interventional distribution $\eta$ used to train. We nonetheless retain this naming scheme since it reflects the fact that one set contains the policy ``interventions'' of interest while the other does not intervene during program execution.}: 
\paragraph{Observationally trained surrogates} The first approach entails training the surrogate models with $\eta$ taken to be a uniform distribution $\mathcal{U}(\intervsetbase_{\rm init})$ over $\intervsetbase_{\rm init}$. 
\paragraph{Interventionally trained surrogates} This second approach entails training with $\eta$ instead taken to be a uniform distribution $\mathcal{U}(\intervsetbase)$ over $\intervsetbase$. 

We indicate the two approaches to training the surrogates with, respectively, bold uppercase ${\bf O}$ and ${\bf I}$. Details of the training procedure and neural network architectures are provided in Appendix \ref{app:exp}. 
We assess the interventional consistency of the surrogates resulting from these two training schemes by computing error metrics on a hold-out test dataset $\mathbf{I}' = \{\bd{y}^{(r')}_{0:T}\}_{r'=1}^{R'}$ of size $R' = 1000$, generated for each $r' \in \intrange{1}{R'}$ as:
\begin{equation}
    \iota^{(r')} \sim \eta = \mathcal{U}\left(\intervsetbase\right),\quad \bd{y}^{(r')}_{0:T} \sim \tau_{\#}\left(\prob_{\scmbase_{\iota^{(r')}}}\right).
\end{equation} 
Specifically, we inspect two error metrics with respect to this hold-out set, for which lower values are better:
\paragraph{The average mean squared error} between trajectories $\tilde{\bd{y}}^{(r')}_{0:T}$ from the surrogates and the $\bd{y}^{(r')}_{0:T}$, i.e.,
\begin{equation}
    {\rm AMSE} = \frac{1}{N^2 R'} \sum_{r'=1}^{R'} \frac{1}{T+1} \sum_{t=0}^T \left\|\tilde{\bd{y}}^{(r')}_t - \bd{y}^{(r')}_t\right\|^2
\end{equation}
where $\tilde{\bd{y}}^{(r')}_{0:T} \sim \prob_{\scmbase^{\psi^\star}_{\omega^{\phi^\star}(\iota^{(r')})}}$ for $r' \in \intrange{1}{R'}$. 

\paragraph{The average negative log-likelihood} of this test data under the likelihood of the learned surrogates, i.e.
\begin{equation}
    {\rm ANLL} = \frac{1}{R'} \sum_{r'=1}^{R'} -\log
    q^{\psi^\star}_{\omega^{\phi^\star}(\iota^{(r')})}(\mathbf{y}^{(r')}_{0:T}),
\end{equation}
Observational consistency is checked on a different hold-out test set $\mathbf{O}'$, generated by taking $\eta = \mathcal{U}(\intervsetbase_{\rm init})$. 

Table \ref{tab:int_obs} shows the values of these performance metrics evaluated on $\mathbf{I}'$ and $\mathbf{O}'$ for all surrogate families and training schemes. We observe that far lower values of the error metrics are obtained by the interventionally, rather than observationally, trained surrogates when assessing interventional consistency. This suggests that training on interventional data can result in more accurate predictions about the effect of interventions in the \abm{}, and that data drawn from the relevant interventional distributions associated with the \abm{} should be included during training if the policy-maker intends to perform policy experiments with the surrogate. Not surprisingly, we also report a minor drop in observational consistency when training with data from the combined intervention set $\intervsetbase$ instead of $\intervsetbase_{\rm init}$, which can be explained by the overfit of the observationally-trained model on the observational test set. We also observe that the \lodernn{} -- which combines the ``mechanistic'' SIRS \ode{} with a flexible \rnn{} -- achieves the best interventional and observational consistencies of the three families of surrogates, suggesting that such hybrid approaches to constructing flexible surrogates are promising choices under our proposed method.

\begin{figure*}[t]
    \centering
    \includegraphics[width=\columnwidth]{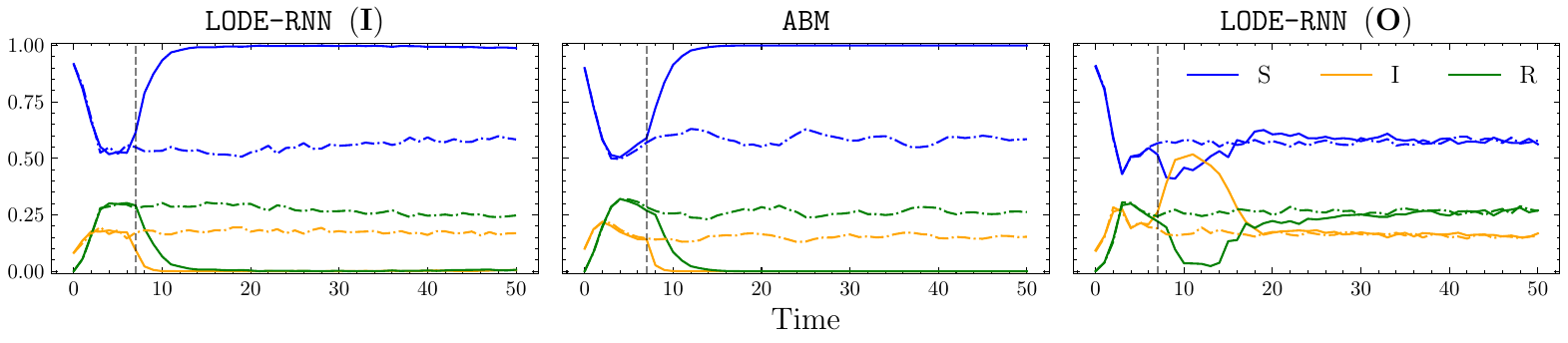}
    \vspace{-0.8em}
    \caption{Example of trajectories from the \abm{} (middle) and the \lodernn{} trained on interventional (left) and observational (right) data. A lockdown is imposed at the dashed vertical line. Solid lines show trajectories under the lockdown, while dot-dash lines show trajectories without lockdown. The effectiveness of the lockdown in reducing infections is vastly underestimated in the observationally trained surrogate, while the interventionally trained surrogate accurately predicts that the lockdown will effectively reduce the spread of infection.}
    \label{fig:lnode_int_obs_example}
\end{figure*}

In Figure \ref{fig:lnode_int_obs_example}, we show an example of a possible negative consequence of failing to train a surrogate on data drawn from the appropriate \abm{} interventional distributions. In the middle panel, we show the change in the \abm{} trajectory induced by imposing a lockdown at time $t_{l} = 7$, while in the left (resp. right) panel we show corresponding trajectories from the interventionally (resp. observationally) trained surrogates under the equivalent intervention learned through our training procedure. While the interventionally trained \lodernn{} correctly predicts that the lockdown effectively impedes the spread of the disease in the \abm{}, the observationally trained surrogate predicts that the lockdown will temporarily \textit{increase} infections, before approximately reverting to the behaviour of the model without a lockdown. The use of such a surrogate model in policy experiments when limited computational resources do not permit use of the accurate, high-fidelity \abm{} of the underlying complex system may therefore have misdirected policy-makers towards suboptimal policies and away from effective interventions. 

\section{Related Work}
Surrogates are frequently used in agent-based modelling.  Increasingly, modern approaches for constructing surrogates rely on established machine learning methods such as random forests \citep{lamperti, airport}, artificial neural networks \citep{epinet, airport}, support vector machines \citep{svm}, kriging \citep{Salle2014}, and mixture density networks (\texttt{MDN}s) \citep{platt2022bayesian}. Our experiments also rely on established machine learning methods to construct surrogates; in particular, our \lrnn{} surrogate family resembles that of \citet{platt2022bayesian}, in which \texttt{MDN}s are used to approximate the transition density for \abm{s}. However, surrogates in agent-based modelling are typically used to expedite simulation-based inference, and their \textit{causal/interventional} consistency with respect to the \abm{} and policy interventions of interest is not generally considered. In contrast to prior work, we explicitly detail the causal structure of our surrogates and their causal relation to the underlying \abm{} via causal abstraction, broadening the scope of surrogate modelling to permit policy experimentation with the surrogate.

The notions of causal abstraction and exact transformation were originally introduced by \cite{rubenstein2017causal}. This seminal work was extended in \cite{beckers2018abstracting}, where stricter definitions of causal abstraction were proposed, and in \cite{beckers2020approximate}, where approximate relationships of abstraction are introduced to account for uncertainty and simplification. The notion of abstraction found practical application in \cite{geiger2021causal} for learning interpretable neural networks.  An alternative category-theoretical definition of abstraction was developed by \cite{rischel2021compositional}, and used for learning abstractions to transfer data between models at different levels of abstraction in \cite{zennaro2023jointly}. 

\section{Conclusion}
Summarising, we have proposed a rigorous framework for learning interventionally consistent surrogates for \abm{s}, formalised through the notion of casual abstraction. Our experiments highlight the efficacy of our framework against purely observational methods that do not train on interventional data. To the best of our knowledge, our framework is the first application of causal abstraction within the \abm{} literature.

In this sense, our work is a first step in formally applying the tools of casual abstraction, and causal inference more generally, to agent-based models. As illustrated in this work, \abm{s} implicitly define causal models for underlying real world systems, and their use in modelling \emph{what-if} scenarios is a causal problem rather than a purely statistical one. Hence, it is natural to conjecture that the theory of causality will play a key role in formalising and addressing many other open problems within the \abm{} community.

\section*{Acknowledgements}

This research was supported by a UKRI AI World Leading Researcher Fellowship awarded to Wooldridge (grant EP/W002949/1).
\textbf{MW} and \textbf{AC} acknowledge funding from Trustworthy AI - Integrating Learning, Optimisation and Reasoning
(TAILOR), a project funded by European Union Horizon2020 research and innovation program under Grant Agreement 952215. 
\textbf{YF}: This scientific paper was supported by the Onassis Foundation - Scholarship ID: F ZR 063-1/2021-2022. \textbf{TD}:
Acknowledges support from a UKRI Turing AI acceleration Fellowship [EP/V02678X/1]. The authors would also
like to acknowledge the University of Warwick Research Technology Platform (RTP) for assistance in the research
described in this paper and the EPSRC platform for ensemble computing ``Sulis'' [EP/T022108/1].


\clearpage
\appendix
\makesupplementtitle
\section{ASSUMPTIONS UNDERLYING MARKOVIAN SCMS}\label{app:assumptions}

Definition \ref{def:scm} implies the standard assumptions of (i) \emph{acyclicity} of the \dirag{} $\scmdag_\scmbase$ and (ii) \emph{causal sufficiency}, meaning that there are no unobserved confounders \citep{pearl2009causality,peters2017elements}. These two assumptions entail that our \scm{s} are Markovian. 

We also assume \emph{faithfulness}, guaranteeing that independencies in the data are captured in the graphical model \cite{spirtes2000causation}.

\section{OTHER NOTIONS OF ABSTRACTION ERROR}
\label{app:details}
As discussed in Section \ref{sec:abserror}, Definition \ref{def:abstractionerr} is closely related to the notion of abstraction error introduced by \cite{beckers2020approximate}. In contrast to Definition \ref{def:abstractionerr}, \cite{beckers2020approximate} employ a maximum over the intervention set $\intervsetbase$ instead of an expectation. Hence, the abstraction error introduced by \cite{beckers2020approximate} may be viewed as a worst-case version of Definition \ref{def:abstractionerr}.

In addition, \cite{beckers2018abstracting} assume the intervention map $\omega$ can be implicitly defined by the map $\tau$, and require the abstraction map $\tau$ to be consistent. That is, the image of $\intervsetbase$ under the intervention map induced by $\tau$ must equal $\intervsetabst$. Since we do not couple the maps $\tau$ and $\omega$ we enforce no such condition. Additionally, \cite{beckers2020approximate} enforce surjectivity of $\tau$. Since this makes no practical difference in a surrogate's use in downstream tasks, we dispense with this assumption.

Alternative notions of abstraction error have been introduced by \cite{zennaro2023quantifying}, building upon the notion of exact transformations introduced by \cite{rischel2020category}. We conjecture that an analogous version of our framework may be developed for this setting, wherein the aggregation function over the intervention set is again chosen to be an expectation over an interventional distribution $\eta$ instead of a maximum, and we leave this as a direction for future work.

\section{ADDITIONAL RELATED WORK}
Surrogate modelling of \abm{s} is closely related to the problem of \abm{} calibration. Calibration involves tuning (a distribution over) \abm{} parameters so that data generated by the \abm{} matches that generated by the real world system being modelled \citep{dyer2022black, dyer2022calibrating}. Analogously, surrogate modelling consists of tuning surrogate parameters so that data generated by the surrogate matches data generated by the corresponding \abm{}. Hence, a number of methods for calibration can naturally be applied to learn surrogates. Several calibration techniques and metrics have been proposed in the literature, including the method of simulated moments \citep{fabretti, gilli, gebbie}, simulated minimum distance \citep{grazzini2015}, generalised subtracted L-divergence \citep{lamperti2018} and Markov information criterion \citep{barde2017}. We refer the reader to \cite{platt, dyer2022a} for thorough surveys. Unlike our framework, these techniques and metrics do not explicitly account for interventional consistency.

More generally, our framework bears many similarities to latent space modelling of Markov decision processes (MDPs) \citep{gelada}, wherein one attempts to learn a smaller latent MDP from a target MDP, whose size precludes its use in downstream tasks. For downstream tasks such as formal verification of policies, \cite{delgrange} employs the bisimulation metric to measure the consistency of their latent MDPs with respect to the target. Abstraction error plays an analogous role in our framework, where the original MDP corresponds to the \abm{}, and the latent MDP the surrogate. Likewise, the surrogates we propose in Section \ref{sec:exp} are implicitly connected to the scientific modelling framework of \cite{rackauckas}, who embed prior information regarding system dynamics into systems of universal differential equations represented by neural architectures such as neural \ode{s}. We embed the underlying dynamics of the classical SIRS \ode{} into several surrogates in an attempt to learn better causal abstractions.

\section{PROOFS}

\subsection{Proof of Proposition \ref{prop:1}}\label{app:proof_1}

\begin{proof}
    By non-negativity of the divergence $d$ we have  $d\left(\tau_{\#} (\prob_{\scminterv{}{\iota}}) ,\, \prob_{\scminterv{\prime}{\omega(\iota)}}\right) \geq 0$ for all $\iota \in \intervsetbase$. Hence  $d_{\tau, \omega}(\scmbase, \scmabst)$ corresponds to an expectation over a non-negative random variable. Since this expectation is equal to zero, we conclude that $d\left(\tau_{\#} (\prob_{\scminterv{}{\iota}}) ,\, \prob_{\scminterv{\prime}{\omega(\iota)}}\right) = 0$ almost surely with respect to the distribution $\eta$. Positivity 
    of the divergence $d$ then implies that $\tau_{\#} (\prob_{\scminterv{}{\iota}})= \prob_{\scminterv{\prime}{\omega(\iota)}}$ almost surely with respect to the distribution $\eta$.
\end{proof}

\subsection{Proof of Proposition \ref{prop:2}}

\begin{proof}
Using Markov's inequality and the fact that $d_{\tau,\omega^{\phi}}(\scmbase, \scmbase^{\psi}) = \mathbb{E}_{\iota \sim \eta}\left[d\left( \tau_{\#}(\prob_{\scmbase_{\iota}}), \prob_{\scmbase_{\omega^{\phi}(\iota)}^{\psi}} \right)\right]$:
\begin{align}\nonumber
    \prob_{\eta}\left( 
    d\left(\tau_{\#} (\prob_{\scmbase_{\iota}})\, \Vert\, \prob_{\scmbase^{\psi}_{\omega^{\phi}(\iota)}}\right)
    \geq 
    \epsilon \right) \leq 
    \frac{ d_{\tau, \omega^{\phi}}\left(\scmbase, \scmbase^{\psi} \right) }{\epsilon}.
\end{align}
Since we have a finite domain, the likelihood functions associated with (a) the pushforward measure of the \abm{} under $\tau$ and (b) the surrogate macromodel can be written as probability mass functions, whose logarithms are non-positive. Since we have assumed $\prob_{\scmbase_{\omega^{\phi}(\iota)}^\psi} \ll \tau_{\#}\left(\prob_{\scmbase_{\iota}}\right)$, we have that $0 \leq -\log q^{\psi}_{\omega^{\phi}(\iota)}(\bd{Y}) < \infty$ for any $\bd{Y} \sim \tau_{\#}\left(\prob_{\scmbase_{\iota}}\right)$, and therefore
\begin{equation}
    0 \leq \mathbb{E}_{\iota \sim \eta}\left[\ce{\iota}\right] < \infty.
\end{equation}
We also have that
\begin{equation}
    -\mathbb{H}_{\tau_{\#}\left(\prob_{\scmbase_{\iota}}\right)} \leq 0\ \Rightarrow\ \mathbb{E}_{\iota \sim \eta}\left[-\mathbb{H}_{\tau_{\#}\left(\prob_{\scmbase_{\iota}}\right)}\right] \leq 0,
\end{equation}
where $\mathbb{H}_{\tau_{\#}\left(\prob_{\scmbase_{\iota}}\right)}$ is the entropy of the probability mass function associated with $\tau_{\#}(\prob_{\scmbase_{\iota}})$, and that
\begin{align}
    &d\left( \tau_{\#}(\prob_{\scmbase_{\iota}}), \prob_{\scmbase_{\omega^{\phi}(\iota)}^{\psi}} \right) = -\mathbb{H}_{\tau_{\#}\left(\prob_{\scmbase_{\iota}}\right)} + \ce{\iota}  \geq 0 \\
    \Rightarrow\ &d_{\tau, \omega^{\phi}}\left(\scmbase, \scmbase^{\psi} \right) = \mathbb{E}_{\iota \sim \eta}\left[-\mathbb{H}_{\tau_{\#}\left(\prob_{\scmbase_{\iota}}\right)}\right] + \mathbb{E}_{\iota \sim \eta}\left[\ce{\iota}\right] \leq \mathbb{E}_{\iota \sim \eta}\left[\ce{\iota}\right].
\end{align}
Hence 
\begin{equation}
    \prob_{\eta}\left( 
        d\left(\tau_{\#} (\prob_{\scmbase_{\iota}})\, \Vert\, \prob_{\scmbase^{\psi}_{\omega^{\phi}(\iota)}}\right)
     \geq 
     \epsilon \right)  \leq  \frac{ \mathbb{E}_{\iota \sim \eta}\left[\ce{\iota}\right] }{\epsilon}.
\end{equation}
\end{proof}

\section{FURTHER EXPERIMENTAL DETAILS}\label{app:exp}

As described in the main text, the three surrogate families we consider have \scm{s} whose corresponding \dirag{s} can be drawn as in Figure \ref{fig:surr_dag}. In this section, we fully specify the corresponding \scm{} for each surrogate. 
Furthermore, for each surrogate, we provide details on the procedure used to train the parameters $\psi$ and $\phi$, which respectively describe the structural equations of each \scm{} and their corresponding intervention map $\omega$.

\subsection{The \lode{} Surrogate Family}\label{app:surrogate}

To construct a set $\scmfamily$ of probabilistic \scm{}s, we define a latent neural ordinary differential equation (\lnode{}) based on the classical SIRS \ode{} system. The SIRS \ode{} system takes the form
\begin{equation}
\begin{aligned}
    \frac{{\rm d} \tilde{S}_t}{{\rm d}t} = \tilde{\gamma}_t \tilde{R}_t - \tilde{\alpha}_t \tilde{I}_t \tilde{S}_t,\ & \quad \quad 
    \frac{{\rm d} \tilde{I}_t}{{\rm d}t} = \tilde{\alpha}_t \tilde{I}_t \tilde{S}_t - \tilde{\beta}_t \tilde{I}_t,\\\label{eq:sirs_ode}
    \frac{{\rm d} \tilde{R}_t}{{\rm d}t} = &\tilde{\beta}_t \tilde{I}_t - \tilde{\gamma}_t \tilde{R}_t,
\end{aligned}
\end{equation}
where $\tilde{\bth}_t = (\tilde{\alpha}_t, \tilde{\beta}_t, \tilde{\gamma}_t) \in \mathbb{R}_{\geq 0}^{3}$ are the \ode{} parameters and 
$\bd{z}_{t} = (\tilde{S}_{t}, \tilde{I}_{t}, \tilde{R}_{t}) \in \mathcal{S}\ \forall t \in [0,T]$ is the \ode{} state, where $\mathcal{S}$ is the two-simplex.   Note that $\bd{z}_{t}$ represents the 
proportion of susceptible, infected and recovered individuals in the population according the SIRS \ode{}. Whilst the parameters $\tilde{\bth}_t$ may change over time -- which will permit the experimenter to intervene on the values of the parameters at different time steps -- we assume the simplest case of assigning the same vector $\tilde{\bd{v}} \in \mathbb{R}_{\geq 0}^{3}$ to all $\tilde{\bth}_t$ when no interventions are applied:
\begin{equation}
    \label{eq:const}
    \tilde{\bth}_{t} = \tilde{\bd{v}}, \quad \forall t \in [0,T].
\end{equation}
In other words, Equation \eqref{eq:const} describes the structural equation $\tilde{f}_{\tilde{\bth}_t}$ for $\tilde{\bth}_t$. Practically speaking, the choice of $\tilde{\bd{v}}$ is inconsequential, as we can model any change to $\tilde{\bth}_{t}$ as an intervention. Given 
$\tilde{\bth}_t$, the \ode{} state $\bd{z}_t$ evolves according to the following rule:
\begin{equation}
    \bd{z}_t = \text{ODESolve}(\bd{z}_{t-1}, \tilde{\bth}_t),\quad t \in \intrange{1}{T},
\end{equation}
where $\text{ODESolve}$ denotes 
numerical integration of 
System \ref{eq:sirs_ode} between times $t-1$ and $t$. In our experiments, we compute this using a Euler scheme with step size $\Delta t = 1$. The initial state of the \ode{} is taken to be $\bd{z}_0 = (1 - \tilde{I}_0, \tilde{I}_0, 0)$. One may change the initial state $\bd{z}_{0}$ through interventions on $\tilde{I}_{0}$, which is modelled as an endogenous variable.

Given $\bd{z}_t$, we draw the endogenous variables $\tilde{\bd{y}}_t$ from a multinomial distribution whose class probabilities are given by $\bd{z}_t$. Whilst $\bd{z}_{t}$ represents the percentage of susceptible, infected, and recovered individuals predicted by the SIR \ode{}, $\tilde{\bd{y}}_t$ represents the actual counts observed by the experimenter. We write $\tilde{f}_{\tilde{\bd{y}}_t}(\tilde{I}_0, \tilde{\bth}_{1:t}, \tilde{\bd{u}}_t)$ to denote the structural function associated with $\tilde{\bd{y}}_t$, where the dependence on $\tilde{I}_0$ and $\tilde{\bth}_{t'}$ for $t' \leq t$ is mediated by the trajectory followed by the $\bd{z}_{t'}$ for $t' \leq t$, and $\tilde{\bd{u}}_t$ are the exogenous random variables required to reparameterise the multinomial sampling procedure on each time step.

Note that $\psi = \emptyset$ for this family of surrogates, and hence $\scmfamily$ is a singleton. For the function $f^{\phi}$ comprising the intervention map $\omega^{\phi}$, we take a feedforward network with layer sizes 3, 32, 64, 64, 64, 32, 3. A ReLU activation is applied after each hidden layer, and a sigmoid activation is applied to the final output layer. The sigmoid activation function ensures that the predicted intervention vector $f^{\phi}(\bd{v})$ on the parameters of the \lode{} has all of its components in the range $[0,1]$, which is suitable when forward simulating the \ode{} with an Euler scheme with $\Delta t = 1$. This feedforward network consists of 12,739 trainable parameters. 

\begin{figure}
\begin{center}
\begin{tikzpicture}[box/.style={draw, text width=1cm,align=center}]
\node[box,  minimum height=1cm, minimum width=2cm] (ode) {ODE \\ Solve};
\node[left=0.5cm of ode.165,font=\bfseries] (th) {$\tilde{\bth}_{t+1}$};
\node[left=0.5cm of ode.195,font=\bfseries] (zt) {$\bd{z}_{t}$};
\node[right=0.5cm of ode.345,font=\bfseries, minimum width=1cm] (zt1) 
{$\bd{z}_{t+1}$};
\node[right=0.5cm of ode.15, font=\bfseries, minimum width=1cm] (ut) {$\tilde{\bd{u}}_{t+1}$};
\node[box,right=2cm of ode, minimum height=1cm, minimum width=2cm] (mn) {MN};
\node[right= 0.5cm of mn, font=\bfseries, minimum width=1cm] (y) {$\tilde{\bd{y}}_{t+1}$};
\draw[->] (th) -- (ode.165);
\draw[->] (zt) -- (ode.195);
\draw[->] (ode.345) -- (zt1);

\draw[->] (ut) -- (mn.165);
\draw[->] (zt1) -- (mn.195);
\draw[->] (mn) -- (y);
\end{tikzpicture}
\end{center}
\vspace{-1em}
\caption{A schematic representation of the \lode{} surrogate family for a single time step. First, the output of the SIRS \ode{} for the next time step, $\bd{z}_{t+1}$, is computed via ODESolve. Then, $\bd{z}_{t+1}$ serves as the logits for a multinomial distribution from which $\tilde{\bd{y}}_{t}$ is sampled. This sampling procedure is denoted by MN in the diagram. The exogenous variables required to reparameterise the multinomial distribution during sampling are denoted by $\tilde{\bd{u}}_{t}$.}\label{fig:lode}
\vspace{-1em}
\end{figure}
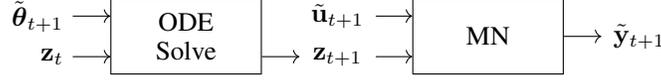

\subsection{The \lodernn{} Surrogate Family}

This surrogate family closely mimics the \lode{} family described above, and differs only in that the class \textit{logits} of the multinomial distributions are instead indexed by the output of a feedforward network -- with layer sizes 32, 32, 64, 32, 16, 3, where all hidden layers are followed by a ReLU activation function -- which maps from the hidden state $\bd{h}_t \in \mathbb{R}^{32}$ of a GRU recurrent network that is passed over the trajectory $\bd{z}_{0:T}$ generated from the SIRS \ode{} (forward simulated as described above). The combined action of the \ode{} solver, the GRU-feedforward networks, and the reparameterisation of sampling from the multinomial distributions, define the structural equations $\tilde{f}_{\tilde{\bd{y}}_t} : (\tilde{I}_0, \tilde{\bth}_{1:t}, \tilde{\bd{u}}_t) \mapsto \tilde{\bd{y}}_t$ for each $t \in \intrange{1}{T}$. 

For this model, $\psi$ is the collection of trainable parameters comprising these GRU and feedforward networks. For $f^{\phi}$, we use a feedforward network with layer sizes 3, 32, 64, 32, 3, where a ReLU activation is applied after all hidden layers and a sigmoid activation is applied after the final layer. Thus, the total number of trainable parameters from $\psi$ and $\phi$ combined is 13,798.

\begin{figure}[t]
\begin{center}
\begin{tikzpicture}[box/.style={draw, text width=1cm,align=center}]
\node[box,  minimum height=1cm, minimum width=2cm] (ode) {ODE \\ Solve};
\node[left=0.5cm of ode.165,font=\bfseries] (th) {$\tilde{\bth}_{t+1}$};
\node[left=0.5cm of ode.195,font=\bfseries] (zt) {$\bd{z}_{t}$};
\node[right=0.5cm of ode.345,font=\bfseries, minimum width=1cm] (zt1) 
{$\bd{z}_{t+1}$};
\node[right=0.5cm of ode.15, font=\bfseries, minimum width=1cm] (ht) {$\bd{h}_{t}$};
\node[box,right=2cm of ode, minimum height=1cm, minimum width=2cm] (rnn) {RNN};
\node[right= 0.5cm of rnn.15, font=\bfseries, minimum width=1cm] (ht1) {$\bd{h}_{t+1}$};
\node[box,right=2cm of rnn, minimum height=1cm, minimum width=2cm] (ff) {FF};
\node[right= 0.5cm of ff.15, font=\bfseries, minimum width=1cm] (ot1) {$\bd{o}_{t+1}$};
\node[right= 0.5cm of ff.345, font=\bfseries, minimum width=1cm] (ut1) {$\tilde{\bd{u}}_{t+1}$};
\node[box,right=2cm of ff, minimum height=1cm, minimum width=2cm] (mn) {MN};
\node[right= 0.5cm of mn, font=\bfseries, minimum width=1cm] (y) {$\tilde{\bd{y}}_{t+1}$};
\draw[->] (th) -- (ode.165);
\draw[->] (zt) -- (ode.195);
\draw[->] (ode.345) -- (zt1);

\draw[->] (ht) -- (rnn.165);
\draw[->] (zt1) -- (rnn.195);
\draw[->] (rnn.15) -- (ht1);
\draw[->] (ht1) -- (ff.165);
\draw[->] (ff.15) -- (ot1);
\draw[->] (ot1) -- (mn.165);
\draw[->] (ut1) -- (mn.195);
\draw[->] (mn) -- (y);
\end{tikzpicture}
\end{center}
\vspace{-1em}
\caption{A schematic representation of the \lodernn{} surrogate family for a single time step. First, the output of the SIRS \ode{} for the next time step, $\bd{z}_{t+1}$, is computed via ODESolve. Then, $\bd{z}_{t+1}$ is passed through to the hidden state of a recurrent neural network (denoted by RNN in the diagram) that updates its hidden state from $\bd{h}_{t}$ to $\bd{h}_{t+1}$. The updated hidden state is passed to a feedforward neural network (denoted by FF in the diagram), which computes the logits $\bd{o}_{t+1}$ for a multinomial distribution from which $\tilde{\bd{y}}_{t+1}$ is sampled.}\label{fig:lodernn}
\vspace{-1em}
\end{figure}
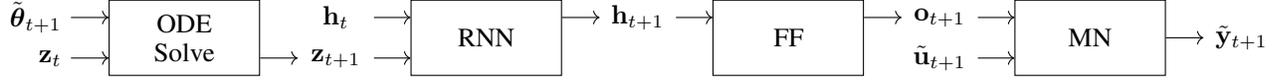

\subsection{The \lrnn{} Surrogate Family}

This surrogate family makes no use of the SIRS \ode{} model. Instead, the logits of the multinomial distributions for $t \in \intrange{1}{T}$ are indexed by the outputs $(\bd{o}_1, \ldots, \bd{o}_T), \bd{o}_t \in \mathbb{R}^3$ of a feedforward network -- with layer sizes 32, 32, 64, 32, 16, 3, and where all hidden layers are followed by a ReLU activation function -- that maps from the hidden state $\bd{h}_t \in \mathbb{R}^{32}$ of a GRU recurrent network which is passed over the sequence $\tilde{\bth}_{1:T}$. The initial hidden state is chosen to be $\bd{h}_0 = (1 - \tilde{I}_0, \tilde{I}_0, \bd{0})$, where $\bd{0}$ is a vector of $30$ zeros. We also take $\bd{o}_0 = (\log(1 - \tilde{I}_0), \log(\tilde{I}_0), -\infty)$ which indexes the logits of the multinomial distribution at time $t = 0$. Once again, we may write the structural equations $\tilde{f}_{\tilde{\bd{y}}_t}$ for the $\tilde{\bd{y}}_t$ in terms of $\tilde{I}_0$, $\tilde{\bth}_{1:t}$, and the exogenous random variables $\tilde{\bd{u}}_t$ required to reparameterise the sampling procedure from the multinomial distribution.  

Since we use exactly the same networks in this surrogate family as in the \lodernn{} family, the total number of trainable parameters from $\psi$ and $\phi$ combined is also 13,798.

\subsection{The likelihood function for each of these surrogate families}

Having intervened on the $\tilde{I}_0$ and $\tilde{\bth}_{t}$ with known values, the class probabilities for each multinomial distribution is completely determined given the deterministic dynamics within the structural equations mapping to the $\tilde{\bd{y}}_t$.

\subsection{Formalising the $\tau$ map}

Taking $\dom{I_0} = \mathcal{J}_\scmbase = [0,1]$, $\dom{\bd{X}_{0:T}} = \mathcal{X}^{T+1}$ with $\mathcal{X} = \{0,1,2\}^{N}$, and $\dom{\boldsymbol{\Theta}_{1:T}} = \mathcal{P}_{\scmbase}^{T}$ with $\mathcal{P} = [0,1]^3$, we define 
\begin{equation*}
\tau : \mathcal{J}_\scmbase \times \mathcal{X}^{T+1} \times \mathcal{P}_{\scmbase}^{T} \to \mathcal{J}_{\scmabst} \times \mathcal{Y}^{T+1} \times \mathcal{P}_{\scmabst}^{T}
\end{equation*}
which operates componentwise as
\begin{equation}
    \tau\left(I_0, \bd{x}_{0:T}, \bth_{0:T}\right) = (\tau_i(I_0), \tau_x(\bd{x}_{0:T}), \tau_{\theta}(\bth_{0:T}))
\end{equation}
where
\begin{align}
    \tau_i &= \id,\\\nonumber
    \tau_x &: \bd{x}_{0:T} \mapsto \left(
    \sum_{n=1}^N \mathbb{I}_{\bd{x}_{nt} = 0},
    \sum_{n=1}^N \mathbb{I}_{\bd{x}_{nt} = 1},
    \sum_{n=1}^N \mathbb{I}_{\bd{x}_{nt} = 2}
    \right)_{0:T},\\
    \tau_\theta &= \id.
\end{align}
In the above, $\id$ is the identity map, and $\tau_x$ acts by counting the total number of susceptible, infected, and recovered individuals in the \abm{} at each time step.

\subsection{Further experimental details on the training procedure}\label{app:exp_training}

All models were trained on a MacBook Pro using an Apple M2 Chip, operating on macOS Ventura 13.2.1. Software dependencies are specified in the GitHub repository containing the code for this paper, which will be made public upon acceptance.

We assume periodic boundary conditions in both spatial dimensions for the \abm{} presented in Example \ref{ex:abm_as_scm}, which is used in all of our experiments.

For the \lode{} and \lodernn{} surrogate families, we forward simulate the SIRS \ode{} with an Euler scheme with step size $\Delta t = 1$. 

For all surrogates, the neural networks comprising the $\omega^{\phi}$ map and structural equations parameterised by $\psi$ were trained with a learning rate of $10^{-2}$ for a maximum number of 1000 epochs, batch size $B = 50$, and with the Adam optimiser \citep{kingma2014adam}. A total number of $R = 1000$ training samples was generated from the \abm{} for each of the observational and interventional training sets; these were each split 5 times into different training and validation sets of sizes $800$ and $200$, respectively, with a new surrogate model trained from scratch on each of these splits. We apply an early stopping criterion in which training is ceased if the validation error does not decrease for 20 consecutive epochs. 

\begin{figure}
\begin{center}
\begin{tikzpicture}[box/.style={draw, text width=1cm,align=center}]
\node[box,  minimum height=1cm, minimum width=2cm] (rnn) {RNN};
\node[left=0.5cm of rnn.165,font=\bfseries] (th) {$\tilde{\bth}_{t+1}$};
\node[left=0.5cm of rnn.195,font=\bfseries] (ht) {$\bd{h}_{t}$};
\node[right=0.5cm of rnn.345,font=\bfseries, minimum width=1cm] (ht1) 
{$\bd{h}_{t+1}$};

\node[box,right=2cm of rnn, minimum height=1cm, minimum width=2cm] (ff) {FF};
\node[right= 0.5cm of ff.345, font=\bfseries, minimum width=1cm] (ot1) {$\bd{o}_{t+1}$};
\node[box,right=2cm of ff, minimum height=1cm, minimum width=2cm] (mn) {MN};
\node[right=0.5cm of ff.15,font=\bfseries, minimum width=1cm] (ut1) 
{$\tilde{\bd{u}}_{t+1}$};
\node[right= 0.5cm of mn, font=\bfseries, minimum width=1cm] (y) {$\tilde{\bd{y}}_{t+1}$};
\draw[->] (th) -- (rnn.165);
\draw[->] (ht) -- (rnn.195);
\draw[->] (rnn.345) -- (ht1);
\draw[->] (ht1) -- (ff.195);
\draw[->] (ff.345) -- (ot1);
\draw[->] (ot1) -- (mn.195);
\draw[->] (ut1) -- (mn.165);
\draw[->] (mn) -- (y);
\end{tikzpicture}
\end{center}
\vspace{-1em}
\caption{A schematic representation of the \lrnn{} surrogate family for a single time step. First, the parameters $\bth_{t+1}$ are passed to a recurrent neural network (denoted by RNN in the diagram) that updates its hidden state. The updated hidden state is passed to a feedforward neural network (denoted by FF in the diagram), which computes the logits $\bd{o}_{t+1}$ for a multinomial distribution from which $\tilde{\bd{y}}_{t+1}$ is sampled.}\label{fig:lrnn}
\vspace{-1em}
\end{figure}
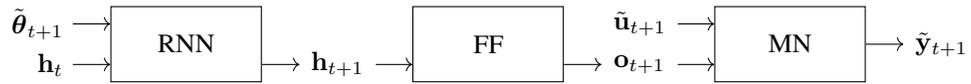



\end{document}